\documentclass[12pt,preprint]{emulateapj}
\newcommand{\chandra}{{\it Chandra}}
\newcommand{\spitzer}{{\it Spitzer}}
\newcommand{\iras}{{\it IRAS}}
\newcommand{\ha}{H$\alpha$}
\newcommand{\hii}{\ion{H}{2}}
\newcommand{\sii}{[\ion{S}{2}]}

\begin{document}

\title{Supernova Remnants in the Magellanic Clouds. VII. Infrared Emission
from SNRs}
\author{R. M. Williams, Y.-H. Chu, and R. Gruendl}
\affil{University of Illinois at Urbana-Champaign, 1002 W. Green St., 
Urbana, IL 61801 USA}
\email{rosanina@astro.uiuc.edu, chu@astro.uiuc.edu, gruendl@astro.uiuc.edu}

\begin{abstract}

We have used the instruments on the \spitzer\ {\it Space Telescope} 
to study the Large Magellanic Cloud supernova remnants (SNRs)  N11L, 
N44, N49, N206, N63A, and N157B.  The two large SNRs N44 and N206 
were not detected in any IRAC or MIPS wavebands; the remainder 
were detected at one or more wavelengths. In particular, the 
SNRs N49 and N63A each had features that were evident in all 
available IRAC and MIPS bands. Each of these two also displayed 
faint limb emission in the MIPS 24 \micron\ band only. IRS 
spectra obtained for the N49 SNR showed a number of prominent 
lines, with little continuum contribution. We therefore suggest
that N49, and possibly N63A, are dominated by line emission, 
with thermal emission from hot dust being at most a secondary
component.

\end{abstract}

\keywords{ISM: supernova remnants  -- ISM: individual (N11L, N44, 
N49, N206, N63A, N157B)}

\section{Introduction}

Supernova remnants (SNRs) are expected to produce infrared (IR) 
emission from such mechanisms as atomic and molecular line emission
(dominated by ground-state fine structure lines), free-free emission 
from hot gas, synchrotron emission, and thermal continuum emission 
from dust heated by collisions with the post-shock gas. Of these, we 
expect only negligible contributions from free-free emission
\citep[e.g.,][]{FTV02} or from  synchrotron emission \citep{G+87a},
except where a pulsar-wind nebula (PWN) may be present.  
Thus, the IR emission is thought to be dominated by line emission 
and/or thermal emission from dust. The dust emission is of particular
interest to the study of SNRs, as supernovae are thought to produce a
significant fraction of dust in the ISM \citep[e.g.,][]{D98}.  Dust 
emission, in turn, is an important factor in cooling the hot plasma 
within SNRs through inelastic collisions with electrons and ions
\citep{OS73,SC01}.

Infrared emission was detected with the {\it Infrared Astronomy
Satellite} (\iras) from roughly a third of the known Galactic SNRs,
demonstrating the difficulty in detecting many of these objects in
the confusion with other IR sources in the ISM. Because of this 
confusion, most of the SNRs to be observed were comparatively 
bright, nearby objects \citep{A89}. Similarly, \iras\ observations 
of the Large Magellanic Cloud (LMC), including pointed observations
toward nine known remnants, only detected four SNRs \citep{G+87a}.
Using the more recent GLIMPSE survey from the {\it Spitzer Space
Telescope}, \citet{R+06} clearly detected 18 of 95 Galactic SNRs 
known to be within the fields.  

\spitzer\ observations are sensitive to many forms of infrared
emission from SNRs, depending on the instrument and wavelength band.  
The Infrared Spectrograph \citep[IRS;][]{Ho+04} is ideal for the study
of line emission.  The Infrared Array Camera 
\citep[IRAC;][]{F+04} is expected to include emission from 
polycyclic aromatic hydrocarbons (PAHs) and very small grains 
(VSGs); H$_2$  lines; and a number of atomic lines including 
Br$\alpha$ 4.1 \micron, [\ion{Fe}{2}] 5.3 \micron\  and [\ion{Ar}{2}]
7.0 \micron.  The Multiband Imaging Photometer for \spitzer\ 
\citep[MIPS;][]{R+04} bands are also expected to include some atomic 
line emission, including [\ion{Fe}{2}] 24.5 \micron, [\ion{O}{1}] 
63.2 \micron, and [\ion{C}{2}] 157 \micron; additionally, we expect
significant thermal emission from collisionally heated dust.

Dust emission from SNRs is of particular interest. It has been 
suggested that supernovae (SNe) are one of the principal sources of 
dust production \citep[e.g.,][]{D98,T+01,D+03}.  Theoretically, models 
have predicted as much as 0.2-4 M$_{\sun}$ of dust production in a 
typical Type II supernova \citep[e.g.,][]{T+01,D+03}.  Certainly 
spectroscopic and photometric observations have suggested dust formation 
in at least five supernovae, including SN1987A \citep{Da05}.  

Finding this freshly produced dust in SNRs, however, has proven 
problematic.   Infrared fluxes attributed to warm ($\sim$ 100-200 K) 
dust have been observed in young remnants such as Cas A, Tycho, and Kepler 
\citep[e.g.,][]{B87,A+99,H+04}, but the inferred dust mass is generally 
several orders of magnitude less than that predicted.  Furthermore, it is 
often unclear whether this dust was largely produced in the SN or has since
been swept up from the surrounding  medium. 

Dust at cooler temperatures ($\le$ 25 K) has also been detected toward
remnants such as Cas A \citep{D+03} and Kepler \citep{M+03} using 
sub-millimeter continuum observations.  Substantially greater dust masses 
($>$ 1 M$_{\sun}$) have been deduced from these observations, but again, 
how much of this dust is ejecta from the recent SN is unknown.  Worse,
not all of the observed dust may be physically associated with the SNRs
themselves.  In fact, \citet{WB05} suggest that ``at least one-half" of 
the ``cold" dust emission from the direction of Cas A is actually 
foreground emission.

\begin{deluxetable*}{lccrrclc}
\tablecaption{Spitzer Datasets used}
\tablehead{
\colhead{SNR} &
\colhead{Inst.} &
\colhead{Mode} &
\colhead{Prgrm} & 
\colhead{AOR} & 
\colhead{PI} &
\colhead{Date} 
}
\startdata
N11L & IRAC & 5$\times$30s Map & 3565 & 11171840 &  Chu & 2004 Nov. 30 \\
N11L & MIPS & Scan Map  & 3565  & 11177728  & Chu  & 2005 Mar. 7 \\
N44-SNR & IRAC & 5$\times$30s Map  & 3565   & 11172352  & Chu  & 2005 Mar. 28 \\
N44-SNR & MIPS & Scan Map  & 3565   & 11177984 & Chu  & 2005 Apr. 7 \\
N49 & IRAC & 10$\times$2s Map  & 124 &  8152064  & Gehrz  & 2004 May 5 \\
N49 & MIPS & 24\micron\ Phot & 124 &   8151808 & Gehrz & 2004 May 26 \\
N49 & MIPS & 70\micron\ Phot & 124 &   8791040 & Gehrz & 2005 May 26 \\
N49 & IRS & Stare & 124 &  6586112  & Gehrz & 2004 May 31 \\
N206-SNR & IRAC   & 3$\times$12s Map & 1061 & 6063104  & Gorjian & 2003 Nov. 21 \\
N206-SNR & MIPS  & Scan Map  & 717 & 7864320 & Rieke & 2003 Nov. 24 \\
N63A & IRAC  & 5$\times$30s Map  & 3565 &  11173888 & Chu & 2004 Dec. 16 \\
N63A & MIPS  & Scan Map & 3565  & 11178496  & Chu & 2005 Mar. 8\\
N157B & IRAC & 3$\times$12s Map  & 63 & 4379904  & Houck  & 2004 Jan. 12\\
N157B & IRAC & 3$\times$12s Map  & 1032 &  6056960  & Brandl  & 2003 Nov. 06 
\enddata
\label{tab:snrobs}
\end{deluxetable*}

\section{Observations}

We obtained proprietary \spitzer\ observations for regions which 
include the LMC SNRs N63A, N11L and SNR 0523$-$67.9 in N44, as 
well as archival \spitzer\ data for SNRs N49, SNR 0532$-$71.0 
in N206, and N157B.  The observations are summarized in 
Table~\ref{tab:snrobs}. All datasets were processed by the \spitzer\  
Science Center (SSC) to produce Basic Calibrated Data (BCD) files.

Images with IRAC  were obtained for all six SNRs. The observations 
were carried out in mapping mode for the 3.6, 4.5, 5.8, and 8.0 
\micron\ bands.  All observations except that for N49 were made
with high dynamic range and cycling dither; the N49 images were
taken with a 5-Gaussian dither.  The exposure times for each map 
are included in Table~\ref{tab:snrobs}.  We used post-BCD mosaic 
images resulting from the standard SSC reduction pipeline (version S11).

MIPS scan maps were available for SNRs  N63A, N11L, N44 (SNR), 
and N206 (SNR) in the 24, 70, and 160 \micron\ bands. All of the
observations were taken at a Medium scan rate, with a scan leg
length of 0\fdg5 and a scan step of 148\arcsec.  Spatial coverages 
were 50\arcmin$\times$20\arcmin\ for  N11L, N44 (SNR), and N206 (SNR),
and 50\arcmin$\times$15\arcmin\ for N63A. For N49, observations in
the MIPS Photometry mode were available in the 24 and 70 \micron\ 
bands. The 24 \micron\ observation took four 3 s observations 
of a small field, while the 70 \micron\ observation took one 
10 s observation of a large field.  We used the standard 
post-BCD images for our study.

N49 was observed with the IRS  in staring mode.  Low-resolution 
spectra were obtained in the 5.2--8.7 \micron\ (SL2) and 7.4--14.5 
\micron\ (SL1) bands; high-resolution spectra were obtained in the 
9.9--19.6 \micron\ (SH) and 18.7--37.2 \micron\ (LH) bands.  For
each instrument, a spectrum is taken at two ``nod" positions, 
intended for on- and off-source comparisons. In this case, both 
positions fall on similar filamentary regions in N49.  IRS 
spectra were extracted and calibrated using the SSC \textsc{SPICE} 
package.  The flux calibration for these spectra is based on SSC
established conversion factors from electrons s$^{-1}$ to Jy, 
corrected by factors determined from the spectra of standard stars. 
No background subtraction was performed, as both ``nod" positions 
lie along emitting material within N49.

\section{Results}

\subsection{Imaging}

There is a great deal of variation as to which SNRs can be seen in 
each of the IRAC and MIPS wavebands (above the level of the nearby 
\hii\ complexes).  Two of these SNRs, N44 and N206, are not detected 
in any band; two others, N11L and N157B, are faintly visible at 4.5 
\micron.  N63A and N49 both have interior concentrations of material 
that are evident at all available wavelengths; in both cases, however, 
the SNR limb is distinct at 24 \micron\ but undetectable in other 
wavebands.

Flux densities for these objects in each waveband are listed in 
Table~\ref{tab:flux}.  When the object is not detected in a given 
waveband, an upper limit for a detection 2$\sigma$ over the background 
is given instead. These fluxes and upper limits are preliminary figures
and should be treated with caution. For one, the SNRs are generally 
superposed on diffuse emission from the nearby \hii\ complexes. To account 
for this and other background emission, we found the median level of 
emission immediately surrounding each SNR, and subtracted this from the 
measured flux values. The uncertainties in the background estimates 
suggest that the listed fluxes are uncertain by 20--30\%.  In addition, 
there are uncertainties in the photometric calibrations of both IRAC 
and MIPS, ranging from at or below 10\% for IRAC and for MIPS 24 \micron, 
to 20\% for MIPS 70 and 160 \micron.  Because these objects are extended, 
the IRAC fluxes are multiplied by the correction factor for an ``infinite" 
aperture as recommended in the IRAC Data 
Handbook\footnote{http://ssc.spitzer.caltech.edu/irac/dh/iracdatahandbook2.0.pdf, 
Table 5.7}: 0.94, 0.94, 0.63 and 0.69 for 3.6, 4.5, 5.8 and 8 
\micron, respectively.  The SNRs are discussed individually below.

\begin{deluxetable*}{lccccccccc}
\tablecaption{Flux Density Estimates}
\tablehead{
\colhead{Band} &
\colhead{N11L} &
\colhead{N44} &
\colhead{N49} &
\colhead{N49-limb} &
\colhead{N206} &
\colhead{N63A} &
\colhead{N63A-limb} &
\colhead{N157B} \\
\colhead{(\micron)} &
\colhead{(mJy)} &
\colhead{(mJy)} &
\colhead{(mJy)} &
\colhead{(mJy)} &
\colhead{(mJy)} &
\colhead{(mJy)} &
\colhead{(mJy)} &
\colhead{(mJy)} 
}
\startdata
3.6 & \nodata & $<$ 74 & 32 & $<$8.8 & $<$ 96 & 37 & $<$ 6.8 & 62 \\ 
4.5 & 1.5 & $<$ 51 & 41 & $<$8.2  & $<$70 & 32 & $<$ 5.9 &  87 \\
5.8 & \nodata & $<$ 86 & 130 & $<$26 & $<$ 140 & 130 & $<$ 17 & $<$ 140 \\
8.0 & $<$ 8.5 & $<$ 190  & 180 & $<$30 & $<$ 250 & 300 & $<$ 45 & $<$ 210\\
24 &  \nodata & $<$ 744 & 1500 & 66 & $<$ 300 & 2300 & 760 & \nodata \\ 
70 &  \nodata & $<$ 5100 & 10,200 & $<$ 433 & $<$ 3200 & 7700 & $<$ 2100 & \nodata \\ 
160 & \nodata & $<$ 12,500 & \nodata & \nodata & $<$ 2200 & 5800 & $<$ 2300 & \nodata \\
\tableline
Area\tablenotemark{a} & 0.73 & 8.80 & 1.11 & 0.40 & 8.31 &  1.31 & 0.92 & 1.20  \\ 
\enddata
\tablenotetext{a}{Area on the sky (of SNR or portion thereof) for 
which flux densities are estimated, given in square arcminutes. The
N49-limb area includes only the eastern limb, without any of the bright
filaments; the N63A-limb area excludes the three-lobed nebular 
emission; and the N157B area excludes the southern dust clouds.}
\tablecomments{``$<$" indicates that the SNR is not detected in this
waveband. The number given is an upper limit, giving the flux required 
for a detection 2$\sigma$ above the background.}
\label{tab:flux}
\end{deluxetable*}

\subsubsection{N11L}

\begin{figure}
\epsscale{1.2}
\plotone{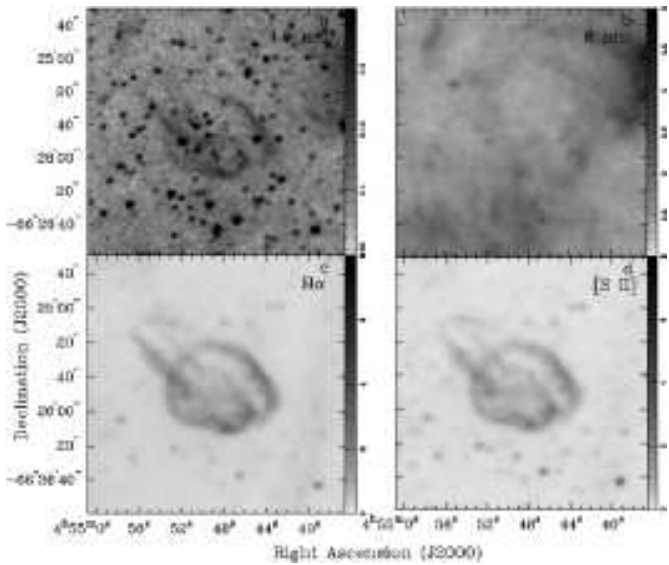}
\caption{Images of N11L with IRAC (a) 4.5 and (b) 8.0 \micron; and
with the Cerro Tololo Inter-American Observatory (CTIO) Curtis Schmidt
Telescope in (c) \ha\ and (d) [\ion{S}{2}].}
\label{fig:n11l}
\end{figure}

N11L (SNR 0454-68.5) is on the periphery of the N11 \hii\ 
complex, framed by a 150 pc filament at the edge of that complex.  
The SNR itself shows a clear shell structure in optical emission 
lines, with filaments projecting beyond the boundary of that
shell. Its optical expansion velocity is $\sim200$ km s$^{-1}$ 
\citep{W+99}.

N11L was close to the edge of the region scanned in the N11 
observation; it was only partially covered in the 3.6 and 
5.8 \micron\  IRAC bands, and was at the extreme edge of the 
three MIPS bands.   No emission from the SNR was distinguished
in any of these bands.  Likewise, the 8 \micron\  IRAC band showed
no emission above the background level for this SNR.  In the 
4.5 \micron\  IRAC band, however, the SNR shell is faint but 
distinct, with a similar morphology to that seen in \ha\
(Fig.~\ref{fig:n11l}).

\subsubsection{SNR 0523-67.9 in N44}

\begin{figure}
\epsscale{1.2}
\plotone{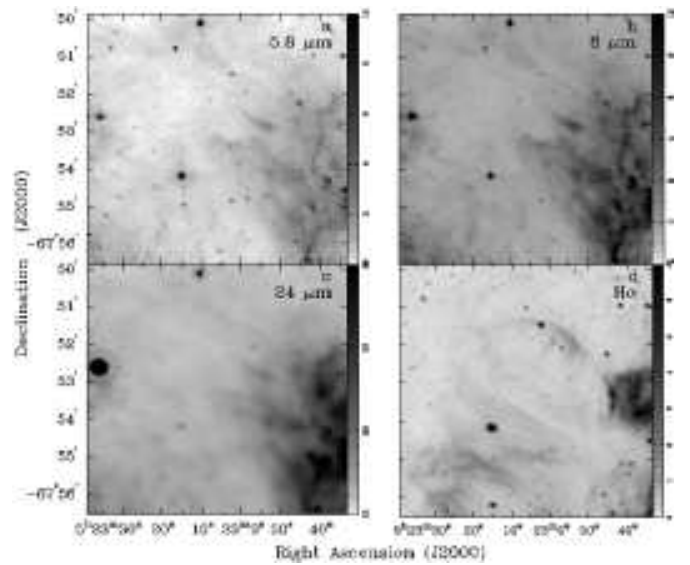}
\caption{Images of N44 with IRAC (a) 5.8 and (b) 8.0 \micron;
MIPS (c) 24 \micron; and the CTIO Curtis Schmidt Telescope in 
 (d) \ha.}
\label{fig:n44}
\end{figure}

The N44 SNR is to the northeast of the N44 \hii\ region, 
overlapping with the \hii\ region to the west and south 
\citep{C+93}.  The SNR has a bi-lobed structure, possibly due
to an intersecting dust lane.  It shows an irregular expansion, 
with velocities of up to 150 km s$^{-1}$ \citep{K+98}.

The SNR in N44 is, in general, conspicuous by its absence in the
IRAC and MIPS images.  There seems to be no infrared emission at 
\spitzer\ wavelengths following the SNR shell seen in the optical 
and X-ray regimes.   A large cloud seen in all bands overlaps 
the SNR southwestern limb; an ``arm" of that cloud seen clearly
at 5.8, 8.0 and 24 \micron\ protrudes across the SNR between its
two lobes.  This ``arm", seen at the IR wavelengths which are most
sensitive to dust emission, is coincident with the dust lane 
causing obscuration in optical images.  Some 5.8 and 8.0 
\micron\ emission surrounds the shell boundary, but aside from 
the intersecting ``arm", the emission in these bands is notably
lower over most of the optical extent of the SNR (Fig.~\ref{fig:n44}). 
While it is quite possible that some of the dust ``arm" is 
shock-heated by the SNR behind it, the ``arm" appears to be 
part of a much larger IR emission feature that extends well 
south of the SNR on the sky.

\subsubsection{N49}

\begin{figure}
\epsscale{1.2}
\plotone{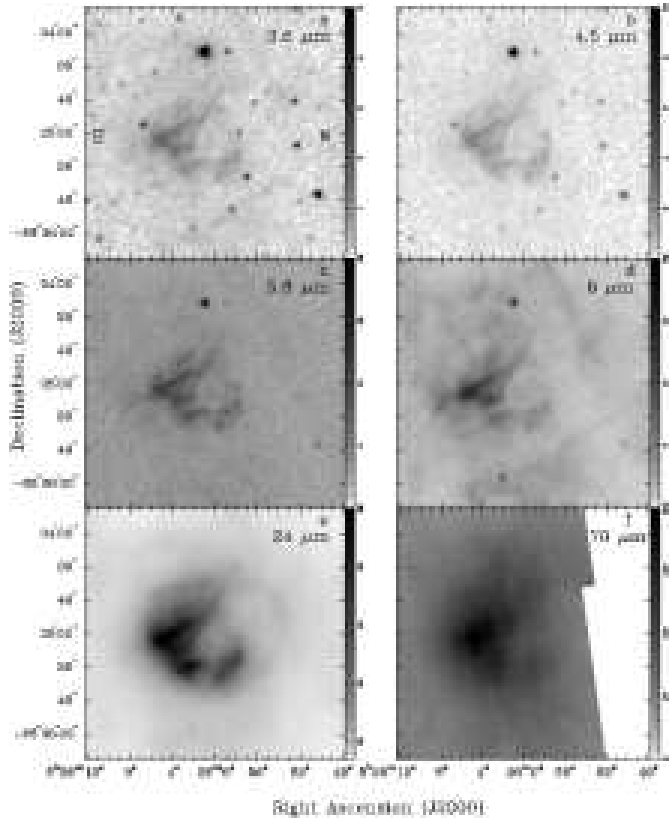}
\caption{Images of N49 with IRAC (a) 3.6 (b) 4.5 
(c) 5.8 (d) 8.0 \micron\ and MIPS (e) 24.0  and (f) 70 \micron. The
E and W in panel (a) label the East and West sides of the SNR.}
\label{fig:n49}
\end{figure}

N49 (SNR 0525-66.1) is unusual for this sample, in that it is 
not situated in or near an \hii\ region.  However, it does lie 
near a molecular cloud, with which it may be interacting 
\citep{B+97}. X-ray and radio maps show a complete SNR shell
\citep{P+03,DM98}, while optical maps largely highlight a 
complex of filaments covering the eastern side of the SNR
\citep{B+06}. It was noted as an \iras\ source by \citet{G+87a}.

\begin{figure}
\epsscale{1.2}
\plotone{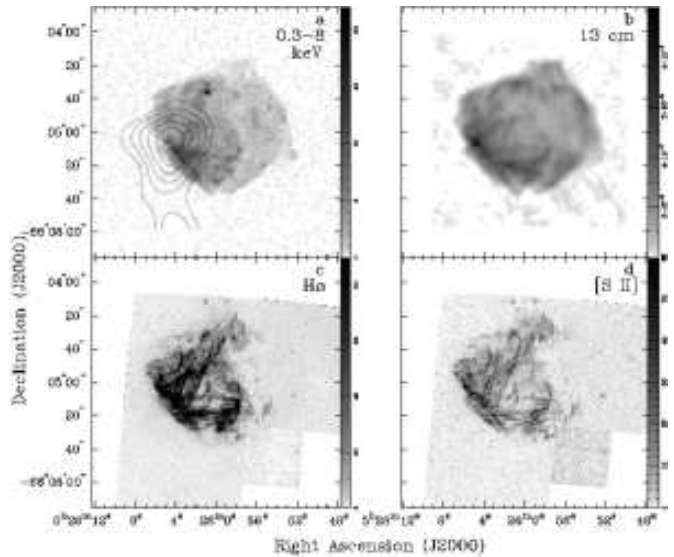}
\caption{Images of N49 with (a) the {\it Chandra} X-ray Observatory
\citep{P+03}; (b) the Australia Telescope Compact Array at 13 cm \citep{DM98}; and the {\it Hubble Space Telescope} (HST) WFPC2 in (c) \ha\ and (d) [\ion{O}{3}] \citep{B+06}. 
The contours in panel (a) are taken from Fig. 5 of \citet{B+97} and show a local CO peak;
 these contours are at 1.0, 1.3, 1.7, 2.0, 2.3 and 2.7 K km s$^{-1}$. }
\label{fig:n49multi}
\end{figure}

Unlike most of the remnants discussed here, N49 shows clearly
in all four of the IRAC bands and the 24 and 70 \micron\ MIPS 
bands (Fig.~\ref{fig:n49}). Observations of this SNR in the 
160 \micron\  band are not yet available in the archive. Regions 
that are bright in the IRAC bands correspond well to brighter 
regions of emission at optical wavelengths \citep{B+06}, even 
showing some of the same filamentary structure 
(Fig.~\ref{fig:n49multi}).  The 70 \micron\ image shows very 
similar structure, albeit at lower resolution.  At 24 \micron,
however, in addition to these bright emission regions, one can 
also detect faint emission from the rest of the 
SNR as seen in the X-ray and radio regimes, with a morphology 
very similar to that in radio.  In particular, the western limb 
of the SNR, which has no detectable optical emission, is seen 
faintly but distinctly at 24 \micron.  In Table~\ref{tab:flux},
we estimate the flux density from the western limb only, as well
as that from the entire SNR.

\begin{deluxetable}{lccccccccc}
\tablecaption{Flux Densities for N63A ``Lobes"}
\tablehead{
\colhead{Band} &
\colhead{N shock} &
\colhead{S shock} &
\colhead{W phot} \\
\colhead{(\micron)} &
\colhead{(mJy)} &
\colhead{(mJy)} &
\colhead{(mJy)} 
}
\startdata
3.6 & 4.6 & 11 & 11 \\ 
4.5 & 4.7 & 9.7 & 9.0 \\
5.8 & 13 & 35 & 30 \\
8.0 & 30 & 93 & 87 \\
24 & 280 & 370 & 370   \\ 
\tableline
Area & 0.044 & 0.054 & 0.036 &  \\ 
\tableline
I(3.6)/I(5.8) & 0.35 & 0.31 & 0.37 \\
I(4.5)/I(8) & 0.16 & 0.10 & 0.10 \\
\enddata
\tablecomments{The three lobes are referred to as ``N shock" for the
northeastern shocked lobe, ``S shock" for the southeastern shocked
lobe, and ``W phot" for the western photoionized lobe. Areas are in
square arcminutes.}
\label{tab:n63flux}
\end{deluxetable}

\subsubsection{N63A}

N63A (SNR 0535-66.0) lies within the N63 \hii\ region \citep{S83}.
It was noted as an \iras\ source by \citet{G+87a}. While radio and
X-ray observations of the SNR show a complete shell \citep{D+93,WHS03},
most of this shell is not seen in optical emission lines. The bright
optical emission is confined to a three-lobed structure on the
western side of the SNR. \citet{L+95} showed that the two eastern 
lobes of this nebula have high \sii/\ha\ ratios, indicative of 
shock ionization, while the optical spectrum of the western lobe 
is more consistent with the photoionization in \hii\ regions.

\begin{figure}
\epsscale{1.2}
\plotone{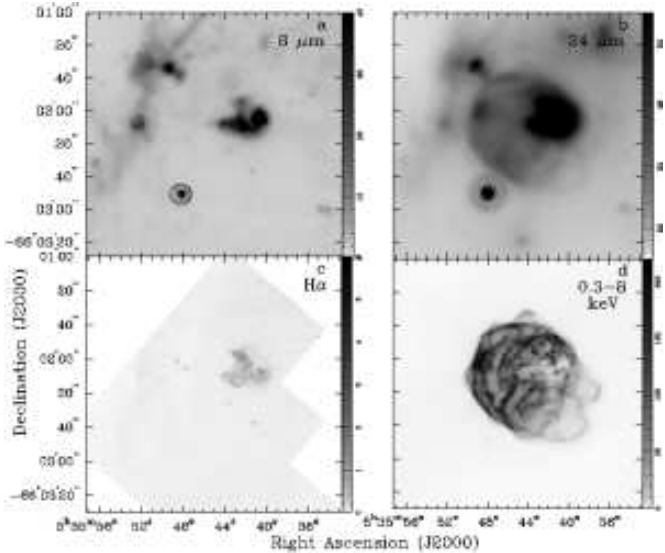}
\caption{Images of N63A with (a) IRAC 8.0 \micron; (b) MIPS 24.0 \micron;
(c) the {\it HST} WFPC2 in \ha; and (d) smoothed X-ray data from 
{\it Chandra} \citep{WHS03}. A background galaxy is marked with a circle 
in (a). }
\label{fig:n63a}
\end{figure}

The optically-bright nebula is also bright in the IRAC and MIPS
wavebands (Fig.~\ref{fig:n63a}).  In the IRAC and 24 \micron\  
MIPS bands, where the resolution is sufficient to distinguish 
between the lobes, the western lobe is notably brighter than the 
two eastern lobes. In the IRAC bands, this three-lobed region is 
the only feature of N63A to appear clearly above the background 
emission.  (A few bright optical knots elsewhere in the SNR 
may have some very faint corresponding emission in the IRAC 
images, but that emission is very close to background levels.)
This also appears to be the case for the 70 and 160 \micron\ 
observations, although the lower resolution of the latter images 
makes it more difficult to distinguish emission associated with 
the three-lobed nebula from the rest of the SNR.  The flux
densities for these lobes are given in Table~\ref{tab:n63flux}.

The 24 \micron\  MIPS band, on the other hand, shows clear 
emission over the entire X-ray/radio shell of the SNR.  The 
emission is evenly distributed over most of the face of the 
SNR, aside from the aforementioned three-lobed structure.  A 
slightly brighter spot on the eastern limb corresponds to an 
area with \sii-bright knots and enhanced X-ray emission 
\citep[Fig.~\ref{fig:n63a}; also][]{C+99}.  
The 24 \micron\ image also shows faint loops to the southwest, 
projecting beyond the boundary of the SNR.  These correspond
well to soft X-ray ``crescents" noted by \citet{WHS03}.  With 
the exception of the three-lobed nebula (seen in X-rays as a
heavily absorbed region), there is a strong correspondence between 
the 24 \micron\ emission and the X-ray emission seen by \chandra.  
The flux density of this 24 \micron\ emission, as with N49, 
is given separately in Table~\ref{tab:flux}, as well as that of 
the SNR as a whole.

\subsubsection{SNR 0532$-$71.0 in N206}

\begin{figure}
\epsscale{1.2}
\plotone{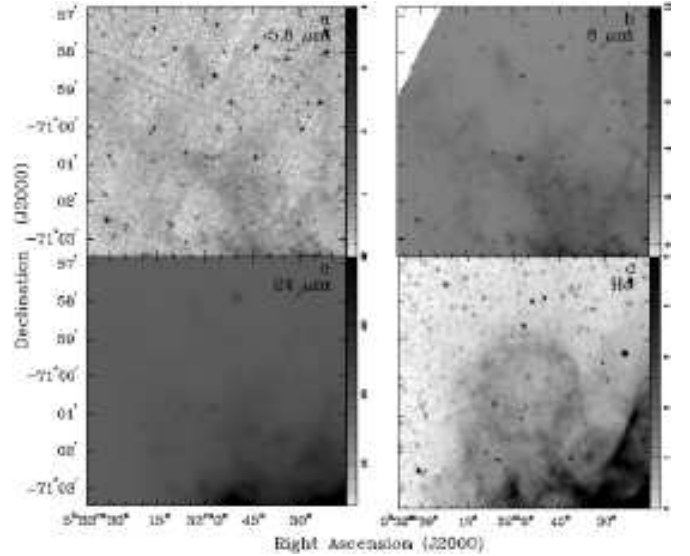}
\caption{Images of N206 with IRAC (a) 5.8 and (b) 8.0 \micron;
MIPS (c) 24 \micron; and the CTIO Curtis Schmidt Telescope in 
 (d) \ha.}
\label{fig:n206}
\end{figure}

The SNR lies just outside the eastern rim of the N206 \hii\ region, 
with some overlap toward the south of the SNR. It was observed 
by \iras\ but not detected \citep{G+87a}.  It has an estimated 
expansion velocity of 200 km s$^{-1}$.  An elongated radio and 
X-ray feature within the remnant, associated with an X-ray point
source, is thought to be a pulsar-wind nebula \citep{W+05}. 
\spitzer\ data for the neighboring \hii\ region have been analyzed 
by \citet{G+04}, who note that the region is rich in PAH emission
{\it except} for the northeast side, nearest the SNR.

As with the SNR in N44, the SNR in N206 does not show emission 
easily distinguishable from that of the surrounding \hii\ region 
(Fig.~\ref{fig:n206}).
An extension from the \hii\ region overlaps the southern limb of
the SNR, but does not seem to be correlated with the emission of 
the SNR at other wavelengths. \spitzer\ images of the N206 \hii\ 
region, including the SNR, have been published by \citet{G+04}.

\subsubsection{N157B}

N157B (SNR 0538-69.1) adjoins the 30 Doradus giant \hii\ region.  
It is one of the two LMC remnants for which the presence of an 
associated pulsar has been directly confirmed. \citet{C+92} used
optical imaging and spectroscopy to determine a maximum SNR extent 
of about 90\arcsec $\times$ 70\arcsec, with patchy optical
emission over its face and no clear shell structure.  Dark 
clouds appear to be obscuring the southern portions of the SNR.

\begin{figure}
\epsscale{1.2}
\plotone{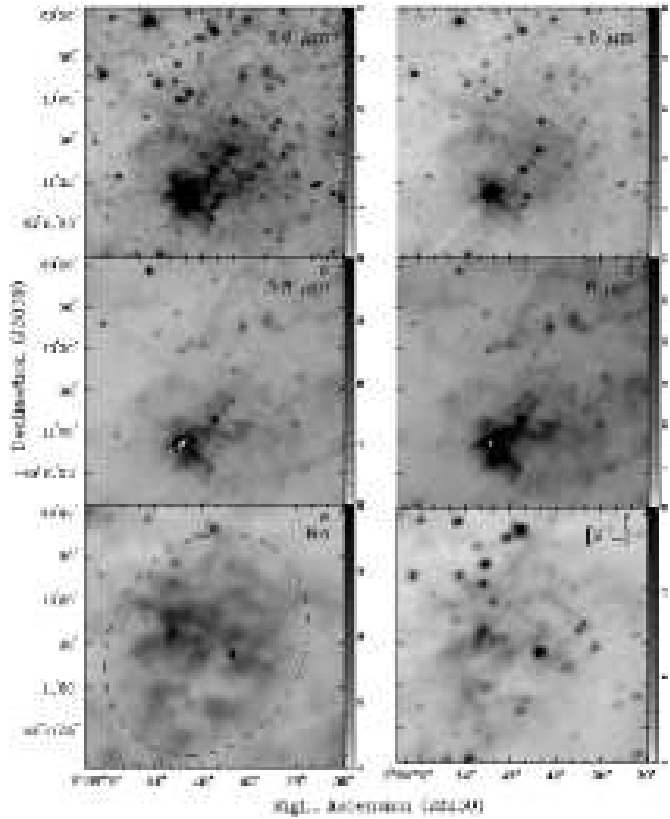}
\caption{Images of N157B with IRAC (a) 3.6 (b) 4.5 (c) 5.8 (d) 8.0 \micron; 
and with the CTIO Curtis Schmidt Telescope in (e) \ha\ and (f) [\ion{S}{2}].
The optical boundary of the SNR is shown by a dotted ellipse in (e).}
\label{fig:n157b}
\end{figure}

No MIPS data are yet available for N157B.  In the IRAC bands
(Fig.~\ref{fig:n157b}), the brightest feature is a ``knot" on the 
southern side of the SNR, with an embedded point source. This bright 
IR region corresponds well to a region of lower \ha\ emission.  
However, there is also fainter emission at 3.6 and 4.5 \micron\ 
north of this dust cloud.  At both these wavelengths, the emission 
is somewhat enhanced (compared to the background) within the 
ellipse that encompasses most of the bright optical emission 
(Fig.~\ref{fig:n157b}e).  In particular, a pair of curved
filaments to the east of the SNR are seen clearly, particularly
at 4.5 \micron.  At 5.8 and 8.0 \micron, any emission from the 
SNR is confused with the extensive foreground/background 
emission in the region.

The faintness of the infrared emission is particularly surprising
when one considers the pulsar-wind nebula within this SNR.  Radio
observations have suggested a spectral index of $\alpha=-0.19$ for
the PWN \citep[$S_{\nu} \propto \nu^{\alpha}$;][]{L+00}.  
Extrapolating from this radio spectrum, we would expect synchrotron
contributions from the PWN of 320$-$360 mJy in the 3.6$-$8 \micron\
range, particularly at the location of the PWN itself as seen in 
radio and X-rays. Instead, the infrared emission, where detectable
above the background, has much lower estimated flux densities and
seems to follow the more prominent optical filaments rather than
the radio/X-ray elongated PWN structure.  

Of course, it is common for PWNe to exhibit a ``broken" power-law 
spectrum, wherein the spectral index steepens as one goes to shorter 
wavelengths.  The spectral break is often attributed to ``aging" by 
synchrotron losses of a population of accelerated particles initially 
distributed with a power-law spectrum.  In N158A (SNR 0540-69.3), the 
other LMC SNR to contain a PWN, this spectral break occurs at ultraviolet 
wavelengths; the 7$-$15 \micron\ fluxes measured from {\it ISOCAM} are 
consistent with an extrapolation from the radio spectral index \citep{GT99}.  
In N158A, also, the spatial distribution of the infrared emission with ISO 
closely matches that of the radio emission.  Recent \spitzer\ observations 
of N158A, however,  suggest that its IRAC-band emission follows the 
extrapolation of a simple power-law from optical wavelengths rather than 
radio -- that is, ``above" the break in the broken power-law spectrum 
\citep{R+06AAS}.  If the spectral break in N157B is similarly located
at wavelengths longer than 8 \micron, the resulting lower flux 
expected from synchrotron emission may be more consistent with our
non-detection.

\subsection{IR Spectroscopy of N49}

\begin{figure}
\epsscale{0.8}
\plotone{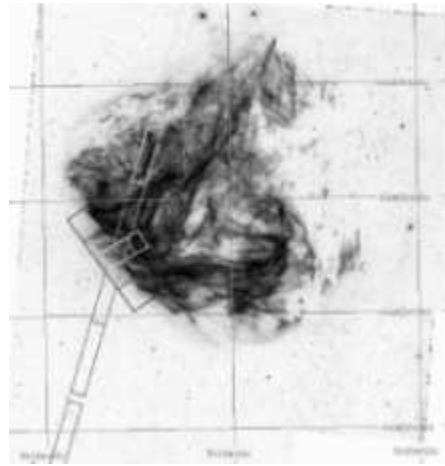}
\caption{Positions of the IRS slits superposed on the {\it Hubble Space Telescope} (HST) WFPC2 \ha\ image of N49.  From shortest to longest,
these are the Short-High Resolution, Long-High Resolution, and Short-Low
Resolution slits.}
\label{fig:slitpos}
\end{figure}

\begin{figure*}
\epsscale{1.0}
\plotone{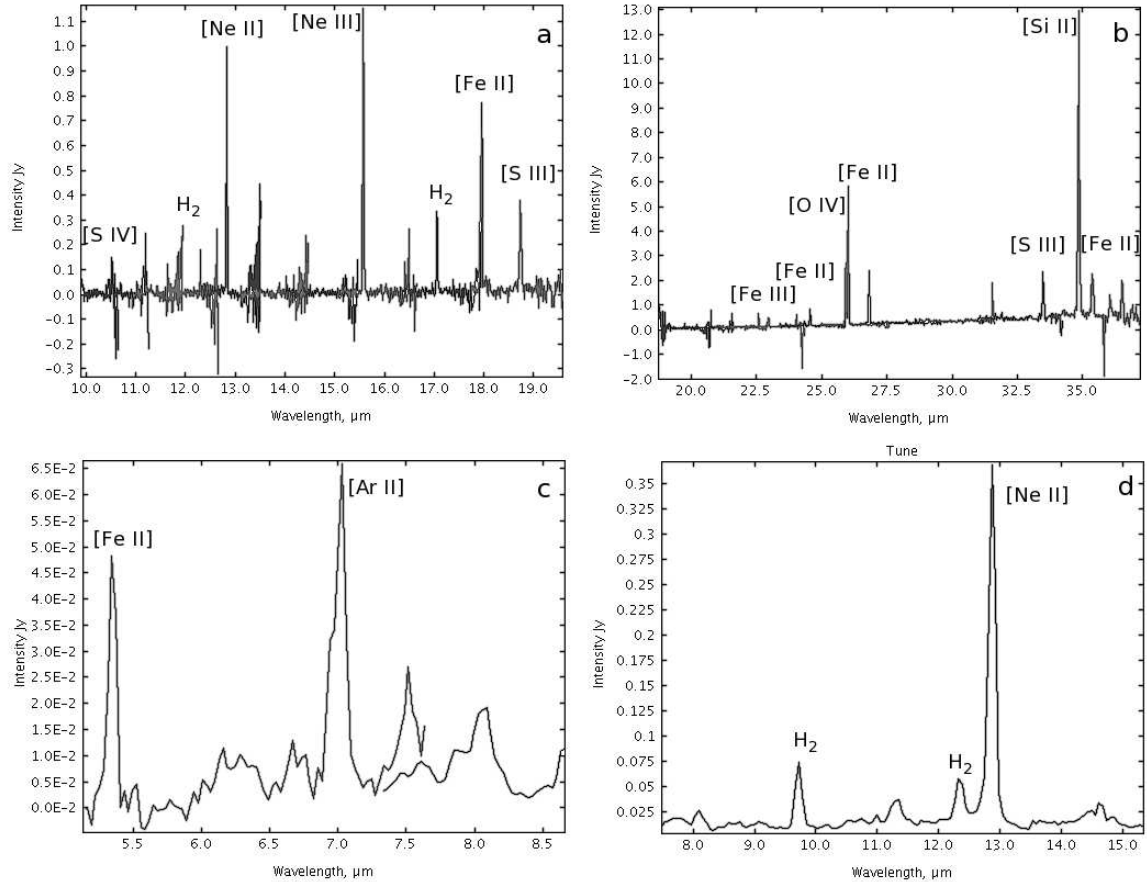}
\caption{Spectra of N49 with IRS: (a) Short-High Resolution, (b) Long-High Resolution, (c) Short-Low Resolution, 2nd-3rd order, and (d) Short-Low Resolution, 1st order.}
\label{fig:n49spec}
\end{figure*}

We examined the available low- and high-resolution IRS spectra for
the SNR N49 at both ``nod" positions. The placement of the slits 
is shown in Fig.~\ref{fig:slitpos}.  Spectra for the first nod
position are shown in Fig.~\ref{fig:n49spec}.  The most prominent 
lines are given in Table~\ref{tab:irslines}, with their measured 
central wavelengths and intensities.  The line intensities are given 
in terms of both peak flux density (Jy) and line flux (erg cm$^{-2}$
s$^{-1}$).   Uncertainties in the flux calibration (for point 
sources) are estimated at 1$\sigma$ values of 20\% for the
low-resolution and 30\% for the high-resolution observations. 
However, the gain along the slit has not yet been fully calibrated, 
so additional errors may be introduced to the flux calibrations for 
our extended sources. There appears to be a significant flux ``jump"
between similar lines observed by the SL and SH modules; in part this
is probably due to the lack of background subtraction, but there is 
also a known mismatch between the detectors\footnote{IRS Data 
Handbook, http://ssc.spitzer.caltech.edu/irs/dh/irsDH10.pdf}.
Tentative identifications are given for each line based on 
wavelength coincidence with ionic and molecular lines known 
to be prominent over the 5.2-38 \micron\ wavelength range in SNRs
\citep[e.g.,][]{O+99,OM+99}.

\begin{deluxetable*}{cccccccccccccccccc}
\tablecaption{Lines and Line Fluxes from IRS data for N49}
\tablehead{
\multicolumn{3}{c}{Nod Position 1} &
\multicolumn{3}{c}{Nod Position 2} &
\colhead{Line} &
\colhead{} \\
\colhead{$\lambda$(\micron )} &
\colhead{F$_{\nu}$\tablenotemark{a}} &
\colhead{F\tablenotemark{b}} &
\colhead{$\lambda$(\micron )} &
\colhead{F$_{\nu}$\tablenotemark{a}} &
\colhead{F\tablenotemark{b}} &
\colhead{ID} & 
\colhead{Comments} 
}
\startdata
\multicolumn{4}{l}{\bf Short--Low Resolution}\\
5.34 & 0.011 & 11  & 5.34 & 0.013 & 8.2 & [\ion{Fe}{2}] &  2nd order \\
7.03 & 0.016 & 11  & 7.00 & 0.027 & 18 & [\ion{Ar}{2}] &  2nd order \\
9.72 & 0.033 & 16  & 9.72 & 0.047 & 19 & H$_2$ & (0,0)S(3) trans \\
12.32 & 0.034 & 14 & 12.32 & 0.041 & 14 & H$_2$&  (0,0)S(2) trans\\
12.87 & 0.189 & 46 & 12.87 & 0.193 & 56 & [\ion{Ne}{2}] & 1st order \\
\multicolumn{4}{l}{\bf Short--High Resolution}\\
10.54 & 0.132 & 3.8  & 10.50 & 0.071 & 0.8 & [\ion{S}{4}] & noise  \\
12.30 & 0.180 & 3.9  & 12.30 & 0.153 & 5.3 & H$_2$ &  (0,0)S(2) trans \\
12.83 & 1.00 & 47  & 12.83 & 0.429  & 20 & [\ion{Ne}{2}] &   \\
15.57 & 1.15 & 43 & 15.57 & 0.567 & 20 & [\ion{Ne}{3}] &   \\
17.05 & 0.333 & 11 & 17.05 & 0.266 & 7.9 & H$_2$ & (0,0)S(1) trans\\
17.94 & 0.665 & 24 & 17.94 & 0.297 & 13 & [\ion{Fe}{2}]  &   \\
18.73 & 0.376 & 14 & 18.73 & 0.183 & 8.6 & [\ion{S}{3}] &   \\
\multicolumn{4}{l}{\bf Long--High Resolution}\\
22.95 & 0.839 & 18 & 22.95 & 0.549  & 12 & [\ion{Fe}{3}]  &   \\
24.55 & 0.247  & 19 & 24.55 & 0.865 & 20 & [\ion{Fe}{2}]  &   \\
25.90 & 2.71 & 55 & 25.90 & 2.75  & 54 & [\ion{O}{4}]  & \\
26.00 & 5.83 & 125 & 26.00 & 6.28 & 115 & [\ion{Fe}{2}]  &   \\
33.47 & 2.37 & 49 & 33.47 & 2.62 & 44 & [\ion{S}{3}]  &   \\
34.84 & 12.9 & 263 & 34.84 & 14.3 & 217 & [\ion{Si}{2}]  &   \\
35.36 & 2.27 & 52 & 35.36 & 2.67 & 46 & [\ion{Fe}{2}]  &   
\enddata
\tablenotetext{a}{\ Peak flux density in Jy} 
\tablenotetext{b}{\ Line total flux in 10$^{-14}$ erg cm$^{-2}$ s$^{-1}$} 
\tablecomments{The ``comments" field includes (a) whether the line is 
seen in first or second order for the Short-Low spectra; (b) which
specific transition of molecular hydrogen causes a given line; and (c) 
whether the line is notably affected by noise in that region of the spectrum.}
\label{tab:irslines}
\end{deluxetable*}

\section{Discussion}

Of our sample of 6 known SNRs within our available fields, we had 
2 strong and 2 marginal detections in at least one waveband,
along with the detection of a feature possibly associated with the 
N44 SNR.  For comparison, 18 Galactic SNRs were strongly detected
(score 1) and 17 possibly detected in confused fields (score 2) by
\citet{R+06}, out of a sample of 95 located within the fields of the
\spitzer\ GLIMPSE survey.  Note, however, that the \citet{R+06} 
search did not include MIPS data, so features which appear 
primarily at 24 \micron, such as the outer limbs of N63A and N49,
would not be seen in their survey.

\subsection{Non-detections}

The non-detections of the SNRs in N206 and, possibly, N44 are
relatively unsurprising. Both of these SNRs have radii $>$20 pc
\citep[Table~\ref{tab:snrs};][]{C+93,W+05}, while the other four 
SNRs have radii of $\sim8-16$ pc. Thus, surface brightness becomes 
more of an issue for these larger remnants, and may well put them 
under the upper limits for detection in most wavebands 
(Table~\ref{tab:flux}).  The lack of detectable emission from the 
N206 SNR is particularly notable in light of the absence of PAH 
emission in that area of the N206 \hii\ region, as noted by \citet{G+04}.

We expect the environments of these SNRs to also have a strong
influence on their detections (or lack thereof).  To show 
substantial dust emission would require the presence of dense
clouds interacting with the SNR.  There is no evidence for 
the presence of such dust clouds in the immediate vicinity of
the N206 SNR; in N44, the optically-obscuring ``arm" suggests 
at least some dense concentrations, but not sufficient for  
shock-heated dust to raise the SNR's IR surface brightness over 
the background.

\subsection{Detections in 1 or 2 bands}

The detection of weak emission from N11L at 4.5 \micron\ and N157B
at 3.6 and 4.5 \micron\ are likely to be due to line emission from
these remnants. Neither SNR is discernable in the dust-sensitive 
8.0 \micron\ band; the shorter-wavelength bands, in general, are
dominated by ionic and (especially) molecular line emission
\citep[e.g.,][]{R+06}.  In N11L, the 4.5 \micron\ morphology closely 
traces the SNR limb and filament structures seen in optical emission 
lines.  In N157B the 3.6 and 4.5 \micron\ emission are enhanced over 
the region where optical filaments are brightest, and in particular 
match the morphology of a pair of bright optical filaments.  The 
relative prominence of the 4.5 \micron\ band, in both cases, suggests 
that Br$\alpha$ 4.1 \micron\ may be a strong contributor to the IR 
emission.

\subsection{Detections in multiple bands}

\subsubsection{Origins of infrared emission in N49}

N49 is bright at all of the wavelengths observed to date by 
\spitzer. This may be a result, in part, from N49's interaction
with a molecular cloud to its southeast.  In this, it is 
similar to the Galactic SNR IC443, which is also interacting 
with a nearby molecular cloud, and is bright at a wide range 
of infrared wavelengths \citep{M+86,BS86,R+01}.   The IRS
observation of N49 confirms the presence of numerous strong 
ionic lines, as well as several lines probably originating from 
H$_2$ transitions. The SNR does not appear to be dominated by 
PAH/VSG emission, as we would expect PAH-dominated emission to 
be strongest in the 8 \micron\ band, while in fact the emission 
there is somewhat closer to background levels than at 3.6 or 4.5 
\micron.  The lack of PAH emission could be accounted for by, e.g., 
PAH destruction by far-ultraviolet (FUV) radiation from the shock
precursor of N49, which is known to have fairly rapid expansion
velocities \citep{B+06}.

We then consider whether thermal emission from dust, as well as 
line emission, makes up a significant part of the IR emission from 
N49.  The emission in the IRAC bands only traces the brightest 
optical filaments, which also show up strongly in the MIPS 24 and
70 \micron\ bands.  These bright filaments most probably are the
source of the emission measured in the IRS spectra, which indicate
a number of strong emission lines.  The continuum levels of the 
post-BCD spectra appear to be very close to zero, strongly suggesting 
a lack of substantial continuum emission from hot dust.  We therefore
argue that line emission is the dominant component of the mid/far 
infrared emission from the bright filaments which are the dominant
source of emission from  N49. While we cannot rule out additional
contributions from hot dust, this lack of continuum emission 
indicates that such contributions are comparatively minor.

Notably, however, the 24 \micron\ emission (only) also shows the 
complete SNR shell of N49, including the west side which is largely 
undetectable in optical emission lines.  The spectroscopy of N49
shows the presence of several lines within the bandwidth of the 
24 \micron\ image ($\Delta\lambda\approx4.7$ \micron), including 
an [\ion{Fe}{3}] line at 23.0 \micron, [\ion{Fe}{2}] lines at 24.5 
and 26.0 \micron, and an [\ion{O}{4}] line at 25.9 \micron.  The 
25.9 and 26.0 \micron\ lines are both considerably brighter than 
any of the lines which fall within the IRAC bands. Thus it is 
plausible that the source of the 24 \micron\ emission from the SNR 
shell may be primarily line emission.  

To quantify this estimate further, we summed the expected fluxes of 
the lines listed above, which, over the 11\farcs1 $\times$ 22\farcs3 
area of the slit, gave a total flux of 2.2$\pm 0.2 \times10^{-12}$ erg 
cm$^{-2}$ s$^{-1}$.  On the MIPS 24 \micron\ image,  we created an 
11\farcs1 $\times$ 22\farcs3 region at approximately the same sky 
location and orientation as the placement of the IRS Long-High slit.
We summed over the surface brightness of this region of the MIPS
image, without background subtraction (since the IRS flux estimates
included both source and background), and obtained a total flux of 
2.8$\pm 0.1\times10^{-12}$ erg cm$^{-2}$ s$^{-1}$.   Thus, based on 
these rough estimates, line emission could be responsible for almost  
80\% of the MIPS 24 \micron\ emission from N49.  Note, however, that 
in addition to the statistical errors given, this estimate is subject 
to uncertainties in the calibrations of MIPS and IRS, particularly 
given the unknown background contribution to the IRS spectra.  It 
is also quite possible that this estimate applies only to the 
comparatively IR-bright eastern region, and cannot be meaningfully
applied to the rest of the limb, particularly the areas seen only
at 24 \micron.

On the other hand, the flux density of thermal emission from hot 
dust is proportional to $Q_{\lambda} B_{\lambda} (T_d)$, where 
$B_{\lambda}$ is the Planck function, $Q_{\lambda}$ is an
efficiency factor for the infrared emissivity, and $T_d$ is the
dust temperature. For metals and crystalline grains in the mid/far-IR,  
$Q_{\lambda} \propto \lambda^{-2}$; typical dust temperatures from a 
young SNR are $\sim 100-130$ K \citep[e.g.,][]{S+05}. Plotting this 
expected curve for the IRAC and MIPS wavebands, we see that the curve 
peaks in the 24 \micron\ band.  The curve predicts that the next highest 
flux density, in the 8 \micron\ band, will be $\sim10\times$ fainter
than that at 24 \micron. Since the 24 \micron\ surface brightness of
the northwest limb of N49 is relatively faint ($\sim4-5\sigma$ above
local background), it is quite plausible that emission ten times
fainter would not be detected.

\begin{figure}
\epsscale{1.0}
\plottwo{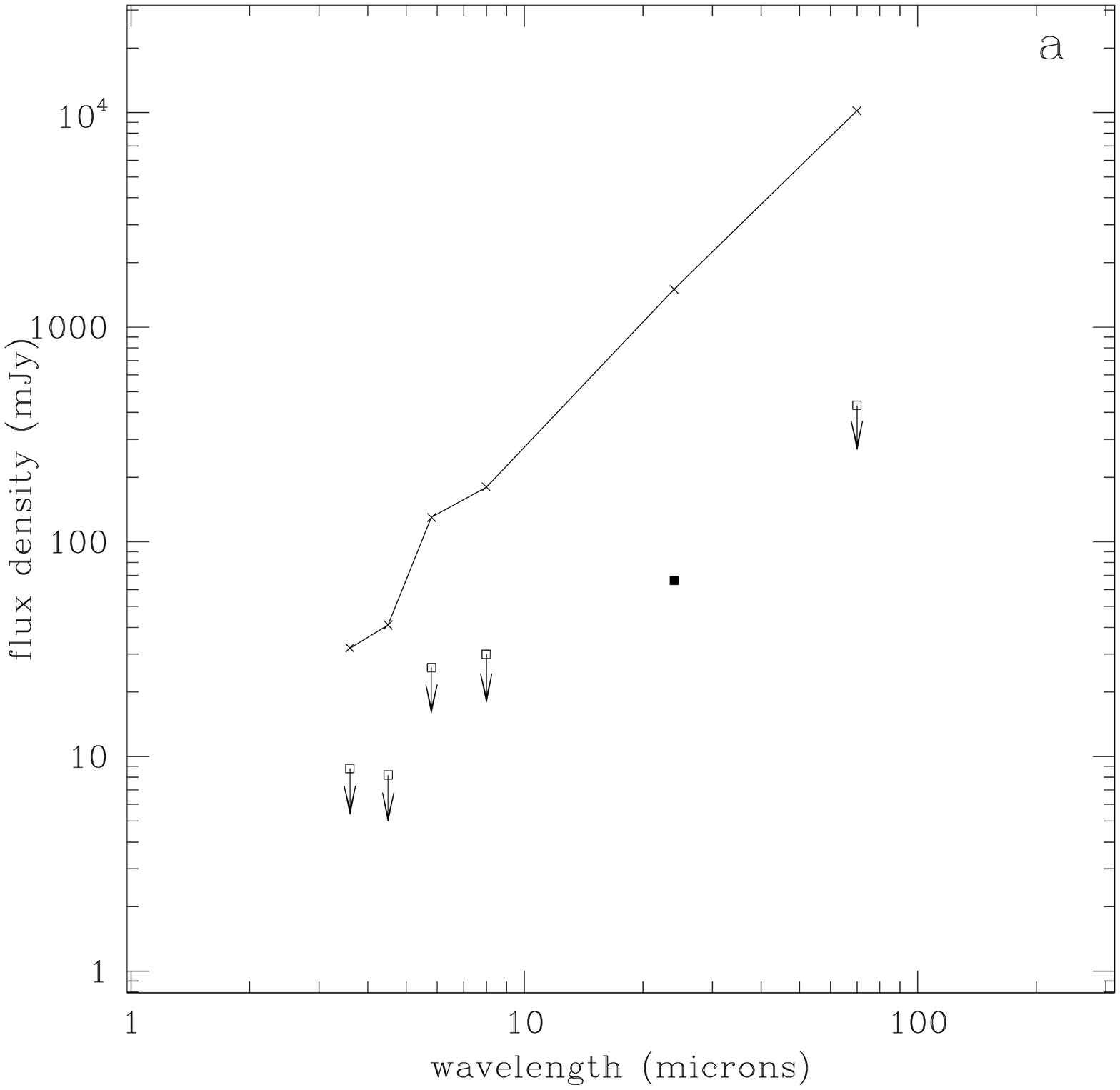}{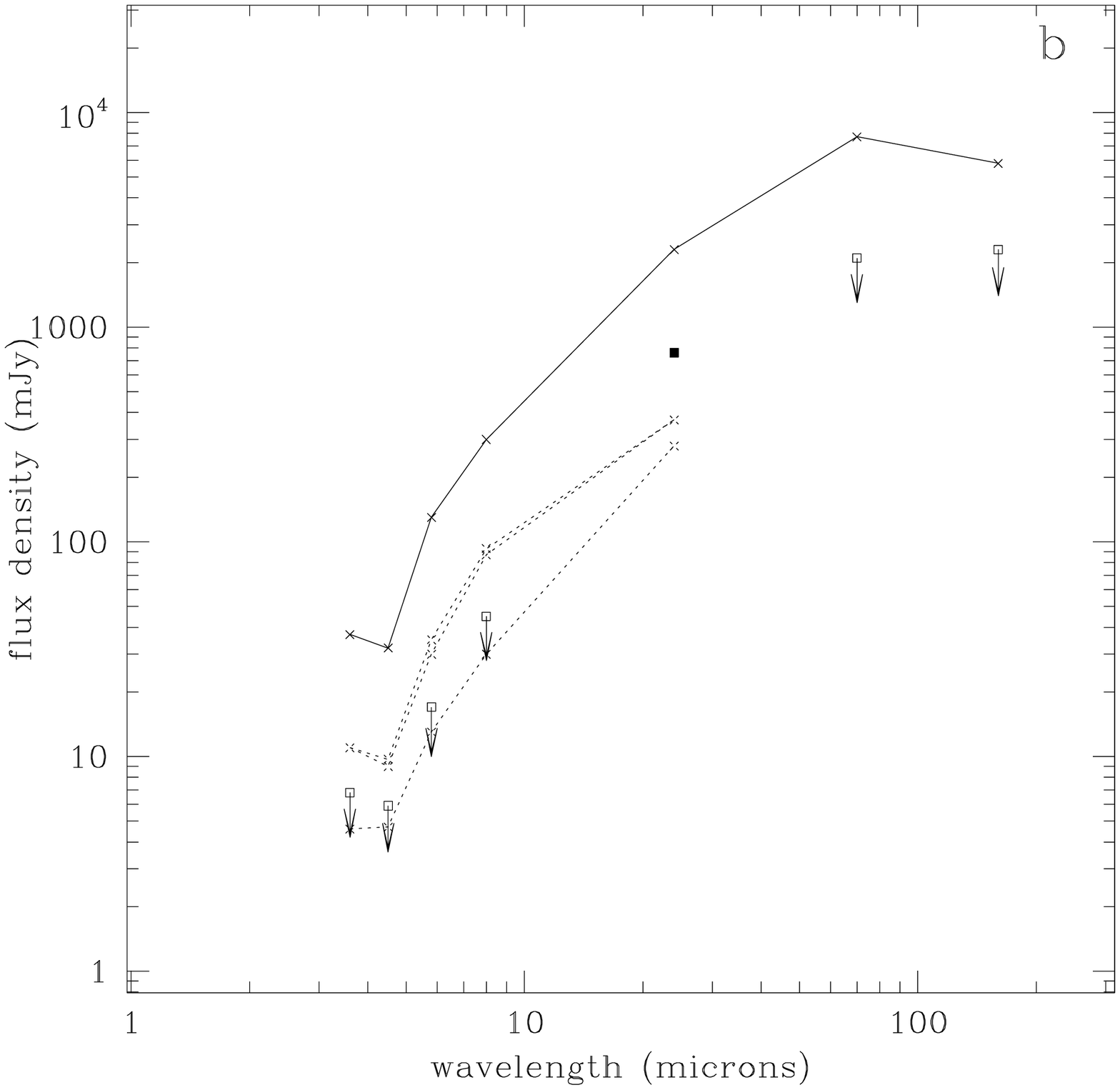}
\caption{Spectral Energy Distributions (SEDS) for (a) N49 and (b) N63A. Solid lines give the SED for the entire SNR;  dotted lines give the SEDs for the three bright clumps in N63A. Filled squares give the 24 \micron\ flux densities along portions of the SNR limb seen only at that wavelength; open squares with arrows give upper limits at other wavelengths.}
\label{fig:seds}
\end{figure}

\subsubsection{Origins of infrared emission in N63A}

 For SNRs within \hii\ regions, such as N63A, there may be sufficient 
far-ultraviolet radiation to destroy nearby PAHs.  In fact, the shell 
(limb) of N63A is largely undetectable in any of the IRAC wavebands, 
which as we have noted above are most sensitive to PAH emission.  At 
24 \micron, however, the complete shell of N63A (also seen in the X-ray 
and radio regimes) is apparent.  The radio and IR coincidence might 
result from a significant synchrotron component in the IR.  However, 
extrapolating the radio spectrum for N63A \citep{D+93} to 24 \micron\ 
gives an expected flux density of 6 mJy at that wavelength, less than
1\% of the observed flux density.  Thus,  we take the strong X-ray and 
IR correspondence as an indication that line emission again dominates 
in the shell at 24 \micron.  As with N49, this does not wholly preclude 
the presence of a hot dust component.  However, the lack of optical 
emission again suggests a low-density environment (other than in the 
three-lobed nebula) and therefore that the region lacks the necessary 
dense clumps for significant dust emission.

In contrast, the three-lobed nebular region of N63A is apparent at 
all available wavelengths, most brightly in the western (photoionized)
lobe.  The spectral energy distribution (SED) of the entire SNR, 
which is dominated by this central nebula, appears similar to that 
of N49.  While it is plausible that these high-density regions in 
N63A may have similar emission properties to the high-density 
filaments of N49, more detailed spectroscopy is required to probe 
the dominant processes producing the infrared emission.  When the 
lobes are considered separately, their SEDs are similar to one 
another overall; surprisingly, the SEDs of the southeastern (shocked) 
lobe and the western (photoionized) lobe have almost identical 
spectral shapes, while the northeastern (shocked) lobe has an SED 
slightly divergent from those of the other two (Fig.~\ref{fig:seds}).
Given that the optical data suggest that the western lobe is 
an unshocked \hii\ region while the eastern lobes are shocked by
the SNR, it is surprising to find that the northeastern lobe,
rather than the western lobe, shows the greatest difference
from the other two lobes.

\begin{deluxetable*}{lccccccccc}
\tablecaption{SNR sizes and ages}
\tablehead{
\colhead{SNR} &
\colhead{Angular} &
\colhead{Linear} &
\colhead{Estimated} &
\colhead{Age} \\
\colhead{Name} &
\colhead{Radii \tablenotemark{a}} &
\colhead{Radii} &
\colhead{Age} &
\colhead{reference} 
}
\startdata
N11L & 35\arcsec $\times$ 30\arcsec & 8.8 $\times$ 7.5 pc  & 7000$-$15,000 yr& \citet{W+99}\\
N44-SNR & 134\arcsec $\times$ 122\arcsec & 33.5 $\times$ 30.5 & $\sim$18,000 yr & \citet{C+93}\\
N49 & 40\arcsec $\times$ 40\arcsec & 10.0 $\times$ 10.0 pc & $\sim$6600 yr & \citet{P+03} \\
N63A & 37\arcsec $\times$ 37\arcsec & 9.3 $\times$ 9.3 pc & 2000$-$5000 yr & \citet{H+98} \\
N206-SNR & 95\arcsec $\times$ 95\arcsec & 23.8 $\times$ 23.8 pc & 23,000$-$27,000 yr & \citet{W+05}\\
N157B & 68\arcsec $\times$ 58\arcsec & 17.0 $\times$ 14.5 pc& $\sim$5000 yr & \citet{WG98} \\
\enddata
\tablenotetext{a}{Size is based on the semimajor and semiminor axis of an 
ellipse fit to the optical extent of the SNR;  except for N49 and N63A, where 
the X-ray extent is used because some of the SNR is not seen in optical.\\ }
\label{tab:snrs}
\end{deluxetable*}

\subsubsection{Infrared and X-ray Comparisons}

One means of addressing the possible contributions of line versus 
continuium emission is to examine the relative distribution of 24 
\micron\ emission compared to the X-ray emission, as done, for example, 
by \citet{S+05} in the study of SNR 1E 0102.2-7219.  Following a similar 
procedure for N49 and N63A, we re-centered and re-gridded the \chandra\ 
X-ray image for each SNR to match the corresponding 24 \micron\ image. 
Each X-ray and 24 \micron\ image was then normalized so that the peak 
emission associated with the SNR was set to 1.0.  (For N49, the X-ray 
emission spatially corresponding to the location of SGR 0525-66 was not 
counted toward this ``peak" emission.)  The resulting normalization 
factors were $7\times10^{-3}$ ct s$^{-1}$ arcsecs$^{-2}$ and 63 MJy 
sr$^{-1}$ for N49, and $5\times10^{-3}$ ct s$^{-1}$ arcsecs$^{-2}$ and 
206 MJy sr$^{-1}$ for N63A.  We then divided the 24 \micron\ images by 
the X-ray images in order to produce ratio maps (not shown), which give 
the distribution of 24 \micron\ emission (with respect to the 24 \micron\ 
peak) compared to the X-ray emission (with respect to the X-ray peak).

If we then consider an ``infrared excess" to be present where these 
ratio maps show ratios greater than 1.0, and if we consider that regions 
dominated by thermal emission from dust should show such an ``infrared 
excess," then we can, by integrating over such regions, estimate a rough 
proportion of the total emission which, we presume, is due to continuum 
emission from dust.  For N49, roughly three-quarters of the flux could 
then be attributed to continuum emission! (Note that the IRS slit 
locations do fall roughly where the ratio maps show an excess of X-ray 
emission compared with IR, so this is not necessarily inconsistent with 
the overwhelming dominance of line emission in the IRS spectra.) For 
N63A, the ``infrared excess" is largely seen on the western side of 
the SNR, including the three-lobed nebula; using the above criteria we 
would expect approximately half the emission to be continuum emission 
from heated dust.

However, this somewhat simplistic approach must be treated with 
caution.  \citet{S+05} used the fact that that the strongest X-ray 
emission in 1E 0102.2-7219 was associated with reverse-shocked ejecta 
in that young SNR, a region which would be expected to generate strong
line emission in many wavelength regimes.  In contrast, N49 and N63A
are both older remnants, in which the ejecta is probably intermixed
with swept-up ISM to a large extent.  For both these SNRs, some 
regions identified as having an ``infrared excess" are also sites of 
significant optical line emission - including parts of the southeastern 
limb of N49 and the ``three-lobed nebula" of N63A.  IR emission 
associated with optically-emitting was suggested by \citet{A+92} to 
include significant infrared line emission!   Other regions, such as
the western side of N49 and the southwestern limb of N63A (the
``X-ray loops"), probably lack significant optical emission 
because that region of the SNR is expanding into a low-density
environment.  This raises the point that we would not expect 
the high-density clouds required to produce substantial thermal 
dust emission in such regions. 

\subsubsection{IRAC colors}

\citet{R+06}, in their study of Galactic SNRs, were able to 
measure fluxes for a number of those SNRs in all four IRAC bands.
Based on measurements and simple models, the authors developed
an IRAC color-color diagram showing the regions of ``color space"
expected for different emission mechanisms, including PAH, ionic,
molecular and synchrotron emission. They found little evidence of
synchrotron emission for their detected SNRs; and state that 
``observations of infrared emission from SNRs to date have shown 
little or no evidence of significant dust emission (PAH or continuum) 
within the wavelengths of the IRAC bands."  From their sample, they
identified regions within the SNRs with colors typical of emission 
from molecular lines, ionic lines, and PAHs.  A number of their 
detected SNRs showed significant molecular-line emission; several
of these have other signs of molecular cloud interaction, such as
maser emission.

N49 and N63A are the only SNRs in this sample for which all four 
IRAC bands are available.  Using  N49,  we determine colors of 
I(3.6\micron)/I(5.8\micron)=0.25 and I(4.5\micron)/I(8\micron)=0.23.
These colors place N49 between the ``molecular shocks" and ``ionic 
shocks" sections of the color-color diagram of \citet[][Fig. 2]{R+06},
somewhat closer to the area where molecular shocks predominate.   
These colors are thought to represent ``a mixture of molecular and
ionic shocks," and are seen in such remnants as, e.g., CTB~37A  
and G348.5$-$0.0.  We see from the ionic and molecular lines
identified by \spitzer\ IRS spectroscopy (Table~\ref{tab:irslines})
that this is indeed an accurate characterization for N49.  
\citet{R+06} notes that both CTB~37A and G348.5$-$0.0 are associated 
with OH 1720 masers, which strongly suggests that these remnants 
are interacting with molecular clouds, as N49 is thought to be.  
All lobes of the three-lobed nebula in N63A have colors 
(Table~\ref{tab:n63flux}) that place them within the ``molecular
shock" portion of the diagram, suggesting that shocked molecular 
gas dominates at the near-IR wavelengths.

\section{Conclusions}

The point that line emission is often a substantial contributor 
to the infrared emission from SNRs has been raised for some time 
\citep[e.g.,][]{G+87b,A+92,O+97,O+99}. For instance, \citet[][]{A+92} 
found that infrared line emission from shocked gas contributes ``a 
significant fraction" to one component of the infrared emission - the 
component generally traced by the presence of optical emission.  (Note 
that dust emission was also thought to contribute to this component, 
and that a second IR component, traced by an excess of IR to X-ray 
emission, was inferred to be generated only by collisionally heated 
dust.)   More recently, \citet{S+05} find that for the Small Magellanic 
Cloud SNR 1E 0102.2-7219, emission from the [\ion{O}{4}] line may 
contribute up to 60\% of the 24 \micron\ emission. 

Our observations of these remnants have primarily shown infrared 
emission, where present, arising from optically-bright areas of the 
SNRs. The implication is that line emission is a significant 
contributor to the infrared emission of these remnants. The
IRAC-band ratios of \citet{R+06} for the optical/IR-bright areas
in N63 and N49 also indicate substantial contributions from ionic 
and molecular lines.  This suggestion is greatly strengthened by 
the infrared spectroscopy of the optically-bright limb of N49, where 
the line contributions appear to overwhelmingly dominate the infrared 
flux density.

Perhaps the most interesting (and mysterious) result is that for
our two clearest detections, N49 and N63A, the outer limb of the 
SNR (in both cases, extremely faint in optical emission lines) 
is detected clearly at 24 \micron\ but not in the other wavebands.
\citet{S+05} find a similar result for 1E 0102.2-7219; the SNR is 
detected clearly at 24 \micron, but not at 8 or 70 \micron.  
Similarly, \citet{Bo+06}, in a study of Type Ia SNRs in the Large 
Magellanic Cloud, find that none of their four SNRs are detected 
in the IRAC bands; two are detected at 24 and 70 \micron, and two 
at 24 \micron\ alone.

The comparative brightness of emission from these SNRs in the 24 
\micron\ band, compared to the other IR bands,  may be due in part 
to the line emission from [\ion{O}{4}] and [\ion{Fe}{2}], particularly 
in regions of bright optical emission.  However, we are far from
able to rule out a secondary dust component even in these regions.
The areas of strong 24 \micron\ emission, weak optical emission, and
moderate X-ray emission may point to a very different physical case, 
where IR emission from collisionally heated dust may yet predominate.
Our future spectral mapping of N49 and N63A with \spitzer's IRS are 
expected to elucidate this situation. 

The potentially significant contributions of infrared line emission 
from at least some SNRs has serious implications for the calculations
of the dust content within those SNRs,  and the use of such 
calculations to constrain models of dust production and destruction.
For example, dust mass estimates based on the 24 \micron\ \spitzer\
flux or the 25 \micron\ \iras\ flux may overestimate the amount of
dust due to overestimation of the flux resulting from inclusion of
line emission.  For some remnants, e.g. where nonradiative shocks
dominate, this difference may be only slight, but for highly 
radiative SNRs, particularly those interacting with dense ambient
material, the effect may be significant.  Likewise, fits to SEDs 
may be unphysical if line contributions in one or more bands are  
not properly taken into account.

\acknowledgements
This work is based on observations made with the \spitzer\ Space 
Telescope, which is operated by the Jet Propulsion Laboratory, 
California Institute of Technology under a contract with NASA. 
Support for this work was provided by NASA through  JPL/Caltech 
award JPL-1264494.  RMW is supported by NASA under grant 
NNG05GC97G through the LTSA program.

\end{document}